\documentstyle[12pt,aasms4]{article}
\lefthead{Park et al.}
\righthead{New Constraints on Simultaneous Optical Emission \\
	From GRBs Measured by the LOTIS Experiment}

\begin{document}

\title{New Constraints on Simultaneous Optical Emission \\
	From GRBs Measured by the LOTIS Experiment}

\author{H. S. Park\altaffilmark{1},
	G. G. Williams\altaffilmark{2},
	E. Ables\altaffilmark{1},
	D. L. Band\altaffilmark{5},
	S. D. Barthelmy\altaffilmark{3,8},
	R. Bionta\altaffilmark{1},
	P. S. Butterworth\altaffilmark{3},
	T. L. Cline\altaffilmark{3},
	D. H. Ferguson\altaffilmark{6},
	G. J. Fishman\altaffilmark{4},
	N. Gehrels\altaffilmark{3},
	D. Hartmann\altaffilmark{2},
	K. Hurley\altaffilmark{7},
	C. Kouveliotou\altaffilmark{4},
	C. A. Meegan\altaffilmark{4},
	L. Ott\altaffilmark{1},
	E. Parker\altaffilmark{1} and
	R. Wurtz\altaffilmark{1}}

% Notice that each of these authors has alternate affiliations, which
% are identified by the \altaffilmark after each name.  The actual alternate
% affiliation information is typeset in footnotes at the bottom of the
% first page, and the text itself is specified in \altaffiltext commands.
% There is a separate \altaffiltext for each alternate affiliation
% indicated above.

\altaffiltext{1}{Lawrence Livermore National Laboratory, Livermore, CA 94550}
\altaffiltext{2}{Dept. of Physics and Astronomy, Clemson University, Clemson, SC 29634-1911}
\altaffiltext{3}{NASA/Goddard Space Flight Center, Greenbelt, MD 20771}
\altaffiltext{4}{NASA/Marshall Space Flight Center, Huntsville, AL 35812}
\altaffiltext{5}{CASS 0424, University of California, San Diego, La Jolla, CA 92093}
\altaffiltext{6}{Dept. of Physics, California State University at Hayward, Hayward, CA 94542}
\altaffiltext{7}{Space Sciences Laboratory, University of California, Berkeley, CA 94720-7450}
\altaffiltext{8}{Universities Space Research Association}

% The abstract environment prints out the receipt and acceptance dates
% if they are relevant for the journal style.  For the aasms style, they
% will print out as horizontal rules for the editorial staff to type
% on, so long as the author does not include \received and \accepted
% commands.  This should not be done, since \received and \accepted dates
% are not known to the author.

\begin{abstract}

LOTIS is a gamma-ray burst optical counterpart search experiment 
located near Lawrence Livermore National Laboratory in California.  
Since operations began in October 1996, LOTIS has responded to five 
triggers as of July 30, 1997, which occurred during good weather 
conditions.  

GRB970223 (BATSE Trigger \#6100) was an exceptionally strong burst 
lasting $\sim30$~s with a peak at $\sim8$ s.  LOTIS began imaging 
the error box $\sim 11$~s after the burst began, and achieved 
simultaneous optical coverage of 100\% of the region enclosed 
by the BATSE $3\sigma$ error circle and the IPN annulus. 
No optical transients were observed brighter than the m$_V \sim 11$
completeness limit of the resulting images providing a new upper limit on 
the simultaneous optical to gamma-ray 
fluence ratio of $R_L < 1.1 \times 10^{-4}$ and on the simultaneous 
optical (at 700 nm) to gamma-ray (at 100 keV) flux density ratio 
of $R_F < 305$ for a B type spectrum and $R_F < 475$ for an M type
spectrum.

\end{abstract}

% The different journals have different requirements for keywords.  The 
% keywords.apj file, found on aas.org in the pubs/aastex-misc directory, 
% contains a list of keywords used with the ApJ and Letters.  These are 
% usually assigned by the editor, but authors may include them in their 
% manuscripts if they wish. 


% That's it for the front matter.  On to the main body of the paper.
% We'll only put in tutorial remarks at the beginning of each section
% so you can see entire sections together.

% In the first two sections, you should notice the use of the LaTeX \cite
% command to identify citations.  The citations are tied to the
% reference list via symbolic KEYs.  We have chosen the first three
% characters of the first author's name plus the last two numeral of the
% year of publication.  The corresponding reference has a \bibitem
% command in the reference list below.
%
% Please see the AASTeX manual for a more complete discussion on how to make
% \cite-\bibitem work for you.   

\section{Introduction}

Gamma-ray bursts (GRBs) are brief bursts of high-energy radiation that 
appear at random positions in the sky.  The origin and nature of GRBs remain
unresolved (\cite{Meegan92}). Despite recent progress, much of the 
difficulty in studying GRBs results from their short duration 
($\sim 1-100$ s) and the poor directional precision 
($\sim 1-10^\circ~1\sigma$ statistical error) available 
from current orbiting GRB  detection experiments (\cite{Fishman95}). 

BATSE (Burst and Transient Source Experiment) on-board the CGRO (Compton
Gamma Ray Observatory) has provided continuous coverage of GRBs for 
the past 6 years. BATSE measures the time histories, spectra, and 
approximate locations of GRBs at a rate of roughly one per day. 
BATSE's most significant observations are that the GRB locations are
isotropic over the sky and that their brightness distribution is 
inhomogeneous. These measurements appear to rule out plausible 
theories of local GRB sources within the plane of our galaxy. 

Recently the Italian-Dutch satellite BeppoSAX (\cite{Costa97a},
\nocite{Costa97b}1997b) has observed two GRBs with 
fading X-ray and optical counterparts 
(\cite{Costa97c}; \nocite{Costa97d}1997d; \cite{Heise97}; 
\cite{vanParadijs97}; \cite{Djorgovski97}).
These counterpart observations were made many hours after 
the GRBs and their association with the GRBs remains to be confirmed.
While these observations provide information on the GRB distance 
scale (\cite{Metzger97}) and their effect on the 
source enviroment, the observed afterglows may result from  
processes different from the prompt gamma-ray production 
mechanism (\cite{Katz97}). Thus, measurements of 
optical emission $\it simultaneous$ with the gamma-ray emission may 
provide crucial clues in understanding this process. 

We  initially adapted an existing wide field-of-view
telescope at Lawrence Livermore National Laboratory (LLNL)
to obtain rapid ($<15$ s) follow-up visual images of GRB 
error boxes utilizing the BATSE real-time coordinate distribution 
network as a trigger (BACODINE/GCN: \cite{Barthelmy94}). This 
instrument, called GROCSE (Gamma Ray Optical Counterpart 
Search Experiment), did not find any evidence for optical 
activity brighter than m$_V \sim  8$ (\cite{Park97}).  

Subsequently, we constructed LOTIS (Livermore Optical Transient Imaging
System), a second generation instrument with 250 times better 
sensitivity than the GROCSE. LOTIS has been operating since October 1996. 
Here we report on the results of LOTIS's search for simultaneous optical 
activity associated with GRBs. 

\section{The LOTIS Experiment}

LOTIS was constructed to respond rapidly to real-time GRB triggers 
provided by GCN (GRB Coordinates Network). The size of the BATSE 
error box, $1\sigma$ error of $\sim 2-10^\circ$, requires wide 
field-of-view optics to obtain 
statistically significant coverage. LOTIS utilizes commercially 
available Canon f/1.8 telephoto lenses, which have short 
200 mm focal lengths and effective apertures of 110 mm diameter.  
The electronic focal plane sensors are $2048 \times 2048$ 
pixel Loral 442A CCDs with $15 \micron \times 15 \micron$ pixels 
driven by custom read-out electronics. The read-out clock rate is 
500 kHz, which results in an image read-out time of 8~s. Each 
Canon telephoto lens/camera assembly has a 
field-of-view of $8.8^\circ \times 8.8^\circ$ with a pixel scale of
15 arcsec.  Four cameras are arranged in a $2 \times 2$ 
array to cover a total field-of-view of $17.4^\circ \times 17.4^\circ$ 
overlapping $0.2^\circ$ in each dimension.

Each of the four cameras has a dedicated SUN/Sparc 2 to control 
image acquisition. The four camera computers are connected to 
the system host computer, a SUN/Sparc 10,  which runs the 
observing program, controls the telescope mount, opens and 
closes the weather protective clamshell housing, reads the 
weather station information, and communicates with GCN.

The on-line software is interrupt-driven and runs 24 hours per day. 
It exchanges ``packets'' (an Internet data exchange protocol) with 
GCN at a rate of one per minute to verify the connection 
between the sites. LOTIS starts the observations every night 
and systematically archives the entire night sky.  
These overlapping fields are saved to tape providing background 
images of future GRB sites. The entire observation program and 
hardware control is fully automated.

Approximately once every 20 days, we receive a ``burst'' packet 
from GCN containing preliminary GRB coordinates. The host 
computer moves the mount rapidly ($<$ 5 s) to the burst coordinates and 
begins imaging the field for an observation period of 20 min. 
During this period approximately 60 images are recorded.

LOTIS is located at Site 300, LLNL's remote test facility, 
25 miles east of Livermore, California. Through July 1997 
LOTIS has accumulated over 1300 hours of background 
images, and five GCN triggers were recorded under good
weather conditions.

We report on one of these triggers, GRB970223 (BATSE Trigger \#6100) 
which occurred on Feburuary 23, 1997 at 8:26:17.67 UTC. 
The high gamma-ray fluence of this event allowed GCN 
to distribute very accurate coordinates, resulting in good 
coverage of its error box.

% Authors may indicate to the editorial staff where they would like 
% figures and tables to be placed in the manuscript.  This is done with
% either the \placefigure{KEY} or \placetable{KEY} commands.  These
% commands require \label{KEY} commands to be placed appropriately with
% corresponding table and figure captions.  When the manuscript is
% printed a short note is printed on the page where the figure or table
% is to go.  These commands are ignored in the aaspp4 and aas2pp4 styles.

%\placetable{tbl-3}
%\placefigure{fig1}

% In this section, we see the use of the \subsection command to set off
% an independent subsection.  We only have one here; usually there would
% be several.

% We show the use of several of the displayed math environments described
% in the User Guide, and you get a healthy dose of mathematical typesetting
% examples.  Also, observe the use of the LaTeX \label command after the
% \subsection to give a symbolic KEY to the subsection for cross-referencing
% in a \ref command.  LaTeX automatically numbers the sections, equations,
% tables, etc., as it goes, so in general you don't know what number something
% is going to have.  We'll refer to the "hairymath" section a little later.

\section{Observation of GRB970223}

GRB970223 had a total gamma-ray fluence of 
$4.8 \times 10^{-5}~{\rm erg~cm^{-2}}$ (20--2000 keV) ranking in the 
top 3\% of all GRBs in intensity.  The gamma-ray intensity 
peaked at $\sim$ 8~s after the start of the burst which lasted 
for $\sim$ 30~s (Figure 1). LOTIS received the GCN burst trigger at 
8:26:23 UTC and began imaging at 8:26:29 UTC, $\sim$ 11~s after the 
burst began. At this time the burst was still in progress, making the 
observation truly simultaneous with the GRB. The shaded 
area in Figure 1 represents the 10~s integration time of the first exposure. 
     
LOTIS imaged the field  centered on the GCN 
coordinates (RA=$144.8^\circ$, Dec=$36.1^\circ$) (J2000.0). 
Later analysis by the BATSE collaboration provided coordinates 
RA=$142.4^\circ$ and Dec=$35.5^\circ$ with $0.73^\circ$ 
statistical ($1\sigma$) and $1.6^\circ$ systematic ($1\sigma$) errors. 
This position was $2.0^\circ$ from the GCN notice but still well 
within the LOTIS field-of-view enabling us to cover 100\% of 
the $3\sigma$ BATSE error box. This event was also observed
with the Ulysses spacecraft (\cite{Hurley97}), allowing an IPN 
(Interplanetary Network) annulus to be constructed, and thus 
further constraining the search region for possible optical counterparts.
Figure 2 shows the BATSE and IPN localizations.

\section{Data Analysis}

The basic data analysis strategy was to search for new star-like  
objects appearing within the GRB error box in the first LOTIS images.  
To determine the exact pointing direction of the LOTIS cameras
during this event, we identified several objects in each image
with stars in the GSC (Guide Star Catalog, \cite{Jenkner90}) and calculated the
elements of the rotation matrix which was used to convert the 
(x,y) pixel positions to celestial coordinates.  Figure~\ref{fig2} 
shows the resulting distribution of the objects detected with LOTIS 
for this event.  There are approximately 3300 star-like objects above 
our $4\sigma$ detection threshold in the full 
field-of-view.  The BATSE $3\sigma$ error circle (including a 
systematic error of 1.6$^\circ$) and the IPN annulus are plotted 
as the ellipse and the narrow arc, respectively.   
Camera 3 has a deficiency of objects because it was 
slightly out of focus during this event.  This did not significantly 
affect our sensitivity since nearly 100\% of the 
error box is covered by the other three cameras.

We used two methods to search for optical transients in the 
first LOTIS images. First, for each object detected with LOTIS within 
the BATSE $3\sigma$ error box and the IPN annulus, we looked for 
corresponding objects in the GSC.  We found that all of the star-like 
objects detected by LOTIS could be identified with a GSC 
object within 4 pixels (60 arcsec).  In the second method, 
we subtracted the final images, taken 20 min after the start of the GRB, 
from the first images, taken 11~s after the start of the GRB and searched 
for new objects at least $4\sigma$ above the background level.  This method 
would reveal faint transient objects present in the first image but not 
the last.  We also examined similar difference images between 
the first images and images of the same field taken 28 days later. 
A few transient objects ($<$ 10) were detected by this method and we 
carefully examined them. They were all either single hot pixels or 
grouping of disconnected single pixels which do not resemble
the point spread function of our system, likely due to cosmic ray hits in
the CCD. All star-like objects present in the first images were also present 
in the later images.  
From these analyses we conclude that no new star-like objects appeared 
in the first 11--21~s after the start of the GRB in the region of sky 
containing the GRB to the sensitivity limits of the LOTIS instrument.

We determined the sensitivity of the LOTIS images by examining 
a histogram of measured object brightness on a log-log scale. 
The measured brightness of each object was determined
in the following way. First, the aperture photometry routines 
in the IDL astronomy package DAOPHOT (\cite{Stetson87}) were 
used to determine an instrumental brightness for 
each object. Then the average ratio of the GSC magnitude 
and the measured instrumental brightness for 20 hand selected stars  was used
to relate the instrumental brightness to visual magnitude for all objects 
in the image.

Figure~\ref{fig3} is the resulting magnitude histogram for the objects
in the first image of cameras 1, 2 and 4.  The peak in the histogram 
measures the completeness limit corresponding to the flux level above which
the entire population of stars was observed (the LogN-LogS 
method).  We find the complete magnitude for this event to 
be m$_{V~complete} = 11.0$. 

\section{Results}
For comparison with models, we derive limits on the ratio 
of optical to gamma-ray fluence and optical to gamma-ray flux density from
our null result.

Since the CCD measures the number of photoelectrons produced
by the objects in the image, our limits are based on the minimum
number of photoelectrons required for detection by our image 
processing algorithms. The number of photoelectrons, $PE_{opt}$, produced 
by a star-like object of magnitude m$_V$ is 
\begin{equation}
PE_{opt} = 2.512^{-m_{V}} \times A_{tel} \times \tau \times 
\int{f(\lambda) \times tr_{atm}(\lambda) \times tr_{tel}(\lambda) 
\times CCD(\lambda)d\lambda}
\label{eq:pe}
\end{equation}
where $A_{tel}$ is the telescope aperture, $\tau$ is the camera 
integration time, $f(\lambda)$ is the stellar spectral flux density in 
${\rm erg ~cm^{-2}~s^{-1}~{\AA}^{-1}}$ at m$_V = 0$ above the 
Earth's atmosphere (\cite{Allen76}), $tr_{atm}(\lambda)$ is the
atmospheric spectral transmission, $tr_{tel}(\lambda)$ is the 
telescope spectral transmission, and $CCD(\lambda)$ is the
CCD quantum efficiency.  To determine the minimum 
photoelectron threshold for detection,  we used 
Equation~\ref{eq:pe} to calculate the absolute expected 
number of photoelectrons produced by the basic
stellar spectral types over the range of magnitudes in 
Figure~\ref{fig3}. We then counted the number of stars in each 
magnitude bin above a trial threshold assuming a distribution of 
stellar spectral types in each magnitude bin 
given by \cite{Hirshfeld82}. This produced a histogram similar 
to Figure~\ref{fig3}.
We varied the trial photoelectron threshold until the 
calculated peak matched 
the measured peak in Figure~\ref{fig3}. The stellar magnitude 
necessary to produce this best fit quantity of photoelectrons 
varied with stellar spectral type from m$_V = 10.4$ for B 
stars to m$_V = 11.3$ for M stars according to Equation~\ref{eq:pe}.

For the first image, the shutter was open from 11 s to 21 s after
the start of the burst. The upper limit to the GRB's optical fluence 
during this time at the top of the atmosphere between 500 nm and 850 nm 
(where our sensitivity is $> 50\%$ of its peak value) is given by 
\begin{equation}
L_{opt} < 2.512^{-m_{V}} \times \tau \times \int_{500 nm}^{850 nm}
{f(\lambda)d\lambda}
\label{eq:lo}
\end{equation}
which yields $L_{opt} < 5.4 \times 10^{-9}~{\rm erg~cm^{-2}}$ 
if the GRB had an optical spectrum of a B type star to 
$L_{opt} < 5.0 \times 10^{-9}~{\rm erg~cm^{-2}}$ if the GRB 
had an optical spectrum of an M type star.

To allow comparisons with models, we normalize this result to the total
gamma-ray fluence from this event. The total GRB fluence during this time 
was calculated by fitting the burst spectrum to the ``GRB'' functional form 
(\cite{Band93}), and then integrating the fit over the 20--2000~keV 
energy range.  BATSE's spectroscopy 
detectors (SDs), specifically SD4, SD6 and SD7, 
were used for the fits. The total GRB fluence for this event 
was $L_{\gamma 20-2000 keV} = 4.8 \times 10^{-5}~{\rm erg~cm^{-2}}$.  
Using these optical and GRB fluences the upper limit of the ratio of the 
optical fluence during our observation to the total gamma-ray fluence is
\begin{equation}
R_{L} = L_{opt: 500 - 850 nm}/L_{\gamma: 20 - 2000 keV} 
< 1.1 \times 10^{-4}
\label{eq:rl}
\end{equation}
for all postulated optical spectra of spectral type B through M.

Finally, since the GRB was still emitting gamma-rays when the 
LOTIS shutters were open, we can present simultaneous limits 
on the ratio of the optical flux at a particular wavelength to the 
gamma-ray flux at 100 keV, the peak of the gamma-ray spectrum. 
We choose 700 nm as the representative wavelength because 
it is at the peak of the spectral acceptance of our detector.
The limiting optical flux density from LOTIS is 
$F_\nu (700~{\rm nm}) < 1.7 \times 10^{-24}
~{\rm erg~cm^{-2}~s^{-1}~Hz^{-1}}$ for GRB optical spectra 
similar to B stars or
$< 2.7 \times 10^{-24}~{\rm erg~cm^{-2}~s^{-1}~Hz^{-1}}$ 
for GRB optical spectra similar to M stars.

The GRB gamma-ray flux density at 100 keV was found by fitting the 
spectra from SD4 and SD6 during the 
LOTIS observation interval ($t=11-21$~s). The 
resulting flux density at 100~keV is 
$F(100~keV)=8.7\times 10^{-3}~{\rm photons~cm^{-2}~s^{-1}~keV^{-1}}$
or
$F(100~keV)=5.8\times 10^{-27}~{\rm erg~cm^{-2}~s^{-1}~Hz^{-1}}$.

The resulting flux ratio limit is 
\begin{equation}
R_{F_{simultaneous}} (t=11-21~{\rm s}) = 
{F_{opt: 700 nm}}/{F_{\gamma: 100 keV}} < 305
\label{eq:rf1}
\end{equation}
for GRB opitcal emission with a B type spectrum and 
\begin{equation}
R_{F_{simultaneous}} (t=11-21~{\rm s}) = 
{F_{opt: 700 nm}}/{F_{\gamma: 100 keV}} < 475
\label{eq:rf2}
\end{equation}
for GRB optical emission with an M type spectrum for this particular burst.

We conclude that any physical mechanisms for GRBs must not yield 
simultaneous optical components larger than these new upper limits.
LOTIS continues to operate and we are now upgrading the 
cameras to enhance our sensitivity to m$_V \sim 16$. Further
constraining limits on simultaneous optical counterparts are expected.

\acknowledgments

This work was supported by the U.S. Department of  Energy, under contract
W-7405-ENG-48 to the Lawrence Livermore National Laboratory and NASA contract
S-57771-F. Gamma-ray burst research at UCSD  (D. Band) is supported by NASA
contract NAS8-36081. K. Hurley acknowledges JPL Contract 958056 for Ulysses
operations and NASA Grant NAG5-1560 for IPN work.

\clearpage

\clearpage

%\figcaption[lotisfig1.eps]{GRB970223 gamma-ray light curve detected 
%by BATSE. The shaded area represents the first LOTIS image exposure 
%showing that we observed this event while the GRB was still in 
%progress. \label{fig1}}

\begin{figure}
\plotone{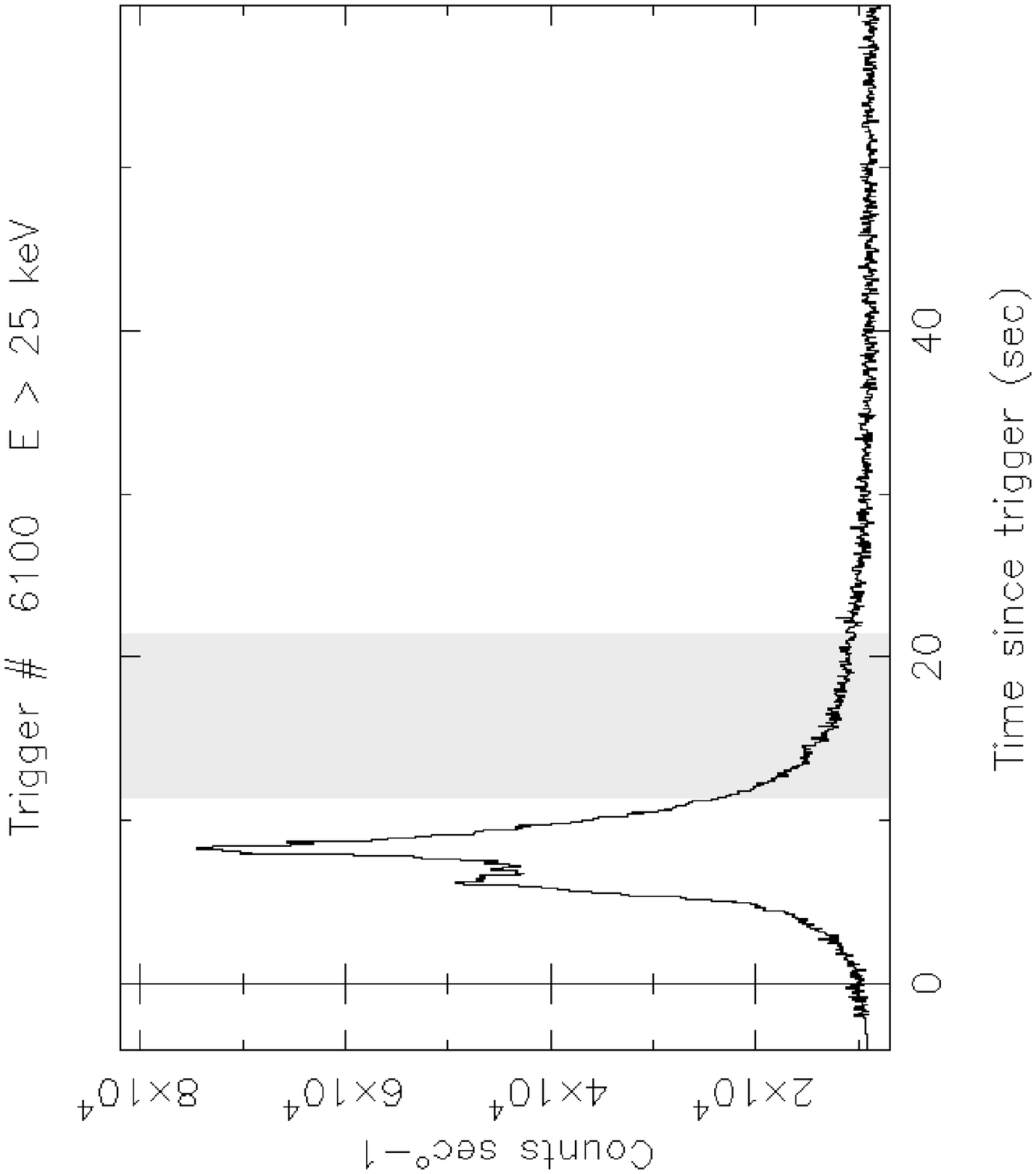}
\caption{GRB970223 gamma-ray light curve detected by BATSE. The shaded area 
represents the first LOTIS image exposure showing that we observed this 
event while the GRB was still in progress. \label{fig1}}
\end{figure}

%\figcaption[lotisfig2.eps]{LOTIS coverage plot for GRB970223. 
%The BATSE 3$\sigma$ error circle and the IPN annulus are plotted 
%as the ellipse and narrow arc, respectively showing that LOTIS 
%covered 100\% of the error box.  The individual dots represent the 
%star-like objects detected above the 4$\sigma$ threshold.\label{fig2}}

\begin{figure}
\plotone{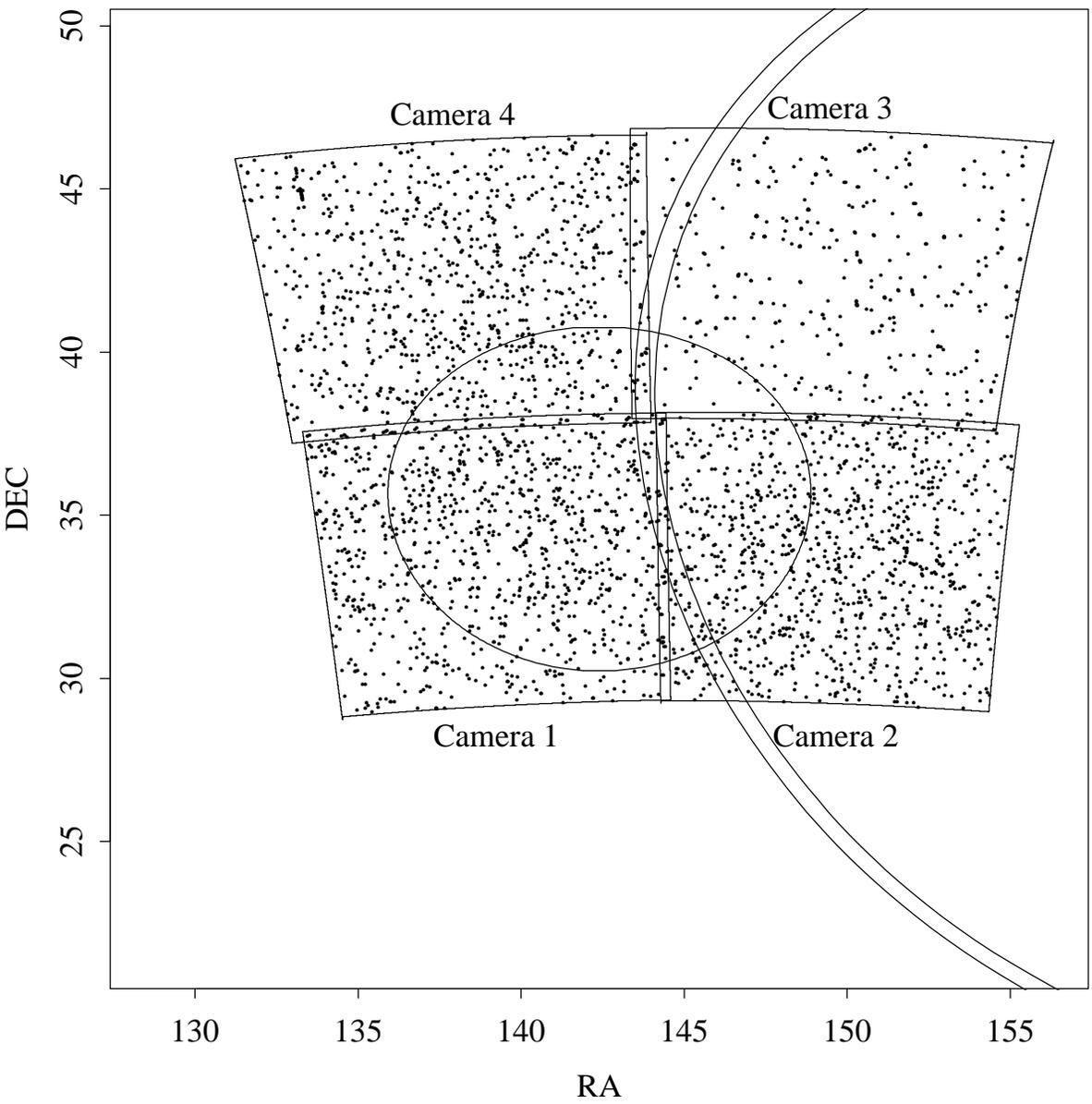}
\caption{LOTIS coverage plot for GRB970223. The BATSE 3$\sigma$ error
circle and the IPN annulus are plotted as the ellipse and narrow arc,
respectively showing that LOTIS covered 100\% of the error box.  The 
individual dots represent the star-like objects detected above the 
4$\sigma$ threshold.\label{fig2}}
\end{figure}

%\figcaption[lotisfig3.eps]{Histogram of stellar magnitudes in the 
%first image of LOTIS camera 1,2 and 4 for GRB970223.  From this 
%we determined a complete magnitude for this event is 
%m$_{V~complete} = 11.0$.\label{fig3}}

\begin{figure}
\plotone{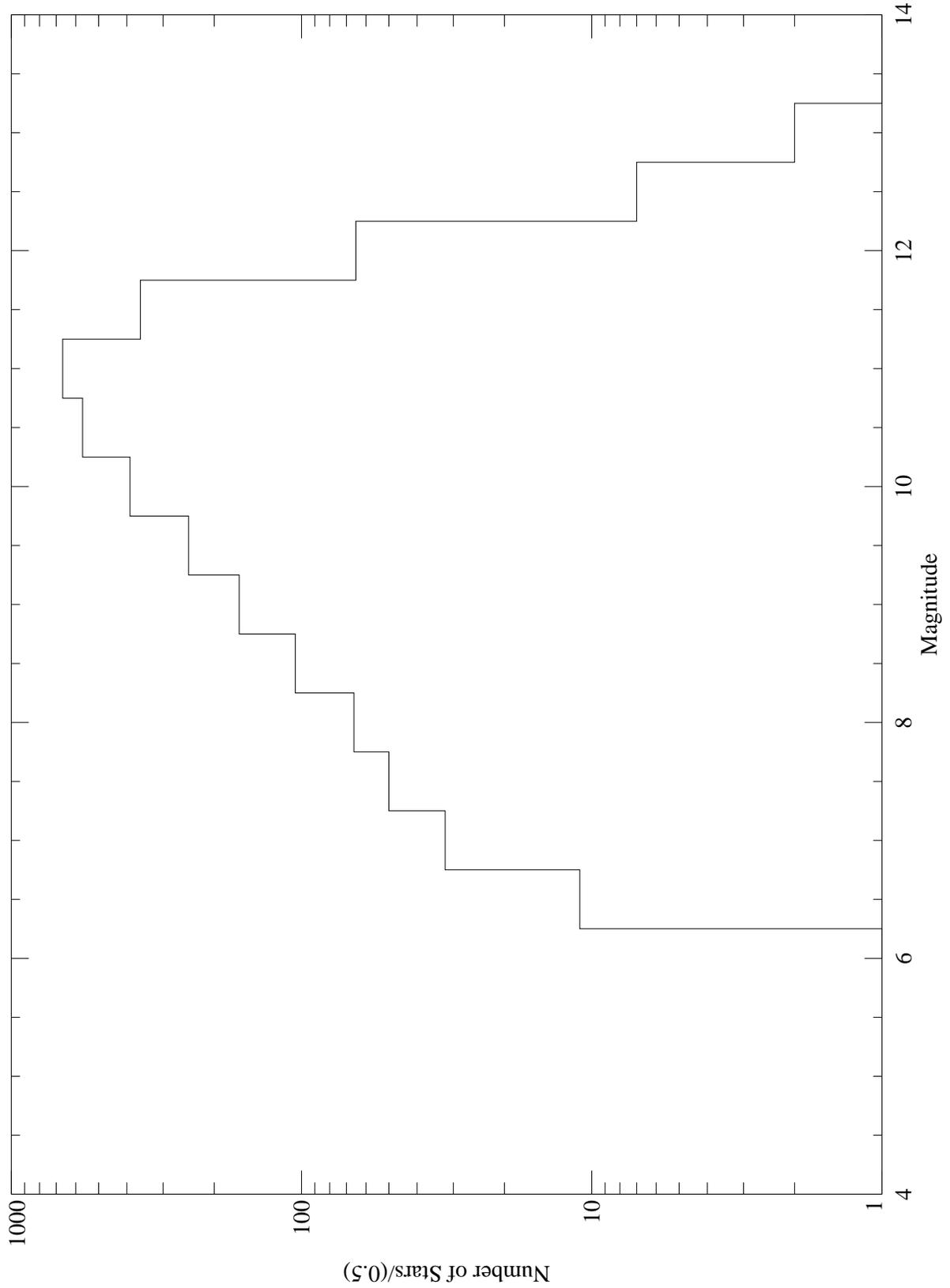}
\caption{Histogram of stellar magnitudes in the first image of
LOTIS camera 1,2 and 4 for GRB970223.  From this we determined 
a complete magnitude for this event is 
m$_{V~complete} = 11.0$.\label{fig3}}
\end{figure}

%\begin{figure}
%\caption{This figure has no associated EPS file, so only the caption
%is printed. \label{fig2}}
%\end{figure}

% The \plotone and \plottwo commands scale the plot(s) in both dimensions
% so that the horizontal dimension fits in the body of the text.  The
% \plotfiddle command will override any automatic scaling, but often
% requires additional "fiddling" to get the plot to fit on the page.
% The \epsscale command allows the author to simply change the scaling
% of the plot in place, without the additional "fiddling" required by 
% \plotfiddle.

%\begin{figure}
%\epsscale{.6}
%\plotone{sgi9259.eps}
%\caption{This is an example of a long figure caption that must be set as
%a paragraph.  The processor has to buffer the text of the
%caption, so it is good not to be too wordy, but that would make for
%poor communication as well. \label{fig3}}
%\end{figure}

% That's all, folks.
%
% The technique of segregating major semantic components of the document
% within "environments" is a very good one, but you as an author have to
% come up with a way of making sure each \begin{whatzit} has a corresponding
% \end{whatzit}.  If you miss one, LaTeX will probably complain a great
% deal during the composition of the document.  Occasionally, you get away
% with it right up to the \end{document}, in which case, you will see
% "\begin{whatzit} ended by \end{document}".

\end{document}